\documentclass[final,1p,times]{elsarticle} 
\usepackage{graphicx} 
\usepackage{amssymb} 
\usepackage{amsthm} 
\usepackage{lineno} 

\journal{Nuclear Physics A} 
\begin{document} 

\begin{frontmatter} 


\title{Measurements of neutral and charged kaon
production at high $p_{T}$ up to 15 GeV/$c$ at STAR}

\author{Yichun Xu \footnote{This work was supported in part by the National Natural Science
Foundation of China  under Grant No. 10610286(10610285), 10475071,
10805046 and the Knowledge Innovation Project of Chinese Academy of
Sciences under Grant No. KJCX2-YW-A14.} for the STAR collaboration}

\address[a]{Department of Modern Physics, University of Science and
Technology of China, Hefei, Anhui 230026, China}

\address[b]{Physics Department, Brookhaven National Laboratory, Upton, NY 11973, USA}
\ead{xuyichun@mail.ustc.edu.cn}

\begin{abstract} 
We report an extension of charged kaon transverse momentum ($p_T$)
spectra at mid-rapidity ($\mid y\mid <$ 0.5) up to 15 GeV/$c$,
neutral kaon $p_T$ spectra up to 12 GeV/$c$ using events triggered
by the Barrel Electro-Magnetic Calorimeter (BEMC) from p+p
collisions at $\sqrt{s_{NN}}$ = 200 GeV. The $K^{\pm}/\pi^{\pm}$ and
$K^{0}/\pi^{\pm}$ at high $p_T$ are compared in p+p and Au+Au
collisions, and nuclear modification factor ($R_{AA}$)for pion,
kaon, proton and rho are discussed. The $R_{AA}$ for kaon in central
collisions are consistent with theory calculation having jet
conversion in a plasma of quarks and gluons.

\end{abstract} 

\end{frontmatter} 



\section{Introduction}
The study of identified hadron ($\pi^{\pm}$, $K^{\pm}$,
$p(\overline{p}$)) spectra at high $p_T$ in p+p collisions provides
a good test of perturbative Quantum Chromodynamics (pQCD)
\cite{pQCD}. In different NLO pQCD calculations, the inclusive
production of single hadron is described by the convolution of
parton distribution functions (PDF), parton interaction
cross-sections and fragmentation functions (FF) which are
parameterized by measured hadron spectra by now. In order to
understand mechanism of hadron production, it is necessary to make a
strict constraint on the quark and gluon FFs by comparing theory
with experimental data. In addition, it's also a good baseline for
studying color charge effect of parton energy loss in heavy ion
collisions, in which hadron spectra are measured up to 12 GeV/$c$
now \cite{starAuAuPID}. Kaon measurements are restricted at high
$p_T$ due to significant uncertainties of ionization energy loss
(dE/dx) from Time Projection Chamber (TPC), while pion and proton
results have been published before \cite{starppPID}. In this
article, we'll present the $p_{T}$ spectra for charged and neutral
kaons in p+p collisions at $\sqrt{s_{NN}}$ = 200 GeV as measured by
the STAR experiment at RHIC, which will be compared with NLO pQCD
calculations. The K/$\pi$ ratios in p+p and Au+Au collisions will be
compared in this paper. Finally, $R_{AA}$, defined by the spectra in
Au+Au collisions divided by spectra in p+p collisions scaled by the
number of binary collisions, are discussed and compared with
predictions from jet conversion \cite{jetConversion}.

\section{Experiment and Analysis}
The data used for this analysis were p+p collisions and Au+Au
collisions collected in the year 2005 and 2004 respectively. The
STAR main tracking detector, TPC covering full azimuthal angle
(2$\pi$) and $\mid\eta\mid < $1.8 in pseudo-rapidity provides a way
to identify charged hadrons by measuring momentum and $dE/dx$
information of charged particles \cite{tpc}. The BEMC \cite{BEMC}
covering 2$\pi$ azimuthal angle and 0$<\eta<$1.0 was used as online
trigger to collect $\sim$5.6 million events with transverse energy
$E_{T} > $ 6.4 GeV (JP2), $\sim$ 5.1 million events with $E_{T}>$
2.5 GeV (HT1), and $\sim$ 3.4 million events with $E_{T}>$ 3.6 GeV
(HT2) from p+p collisions in year 2005. Those trigger can improve
the possibility of events with high momentum track, and this helps
us to extend our measurements up to high $p_T$. JP2 triggered events
are used to identify charged kaons, while HT1 and HT2 triggered
events are used to reconstructed neutral kaons through $K^{0}_{S}
\rightarrow \pi^{+} + \pi^{-}$ decay mode, which are used to cross
check if there are trigger bias by comparing $K^{0}_{S}$ in
different triggered events and $K^{0}_{S}$ with $K^{\pm}$. In
addition, $\sim$ 21.2 million events from central Au+Au collisions
were used to reconstruct neutral kaons and identify charged kaons.


\begin{figure}[ht]
\centering
\includegraphics[scale=0.3]{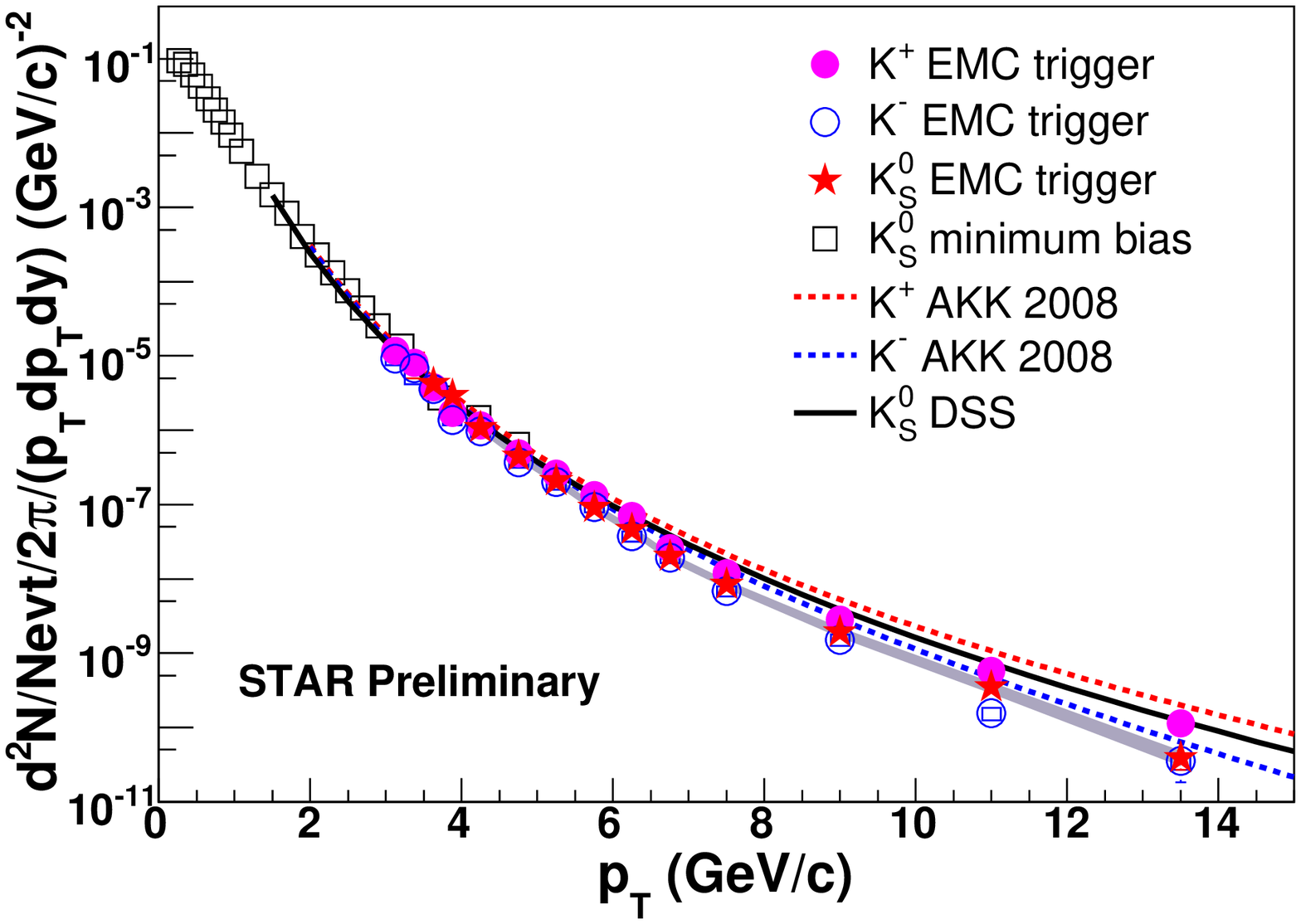}
\includegraphics[scale=0.3]{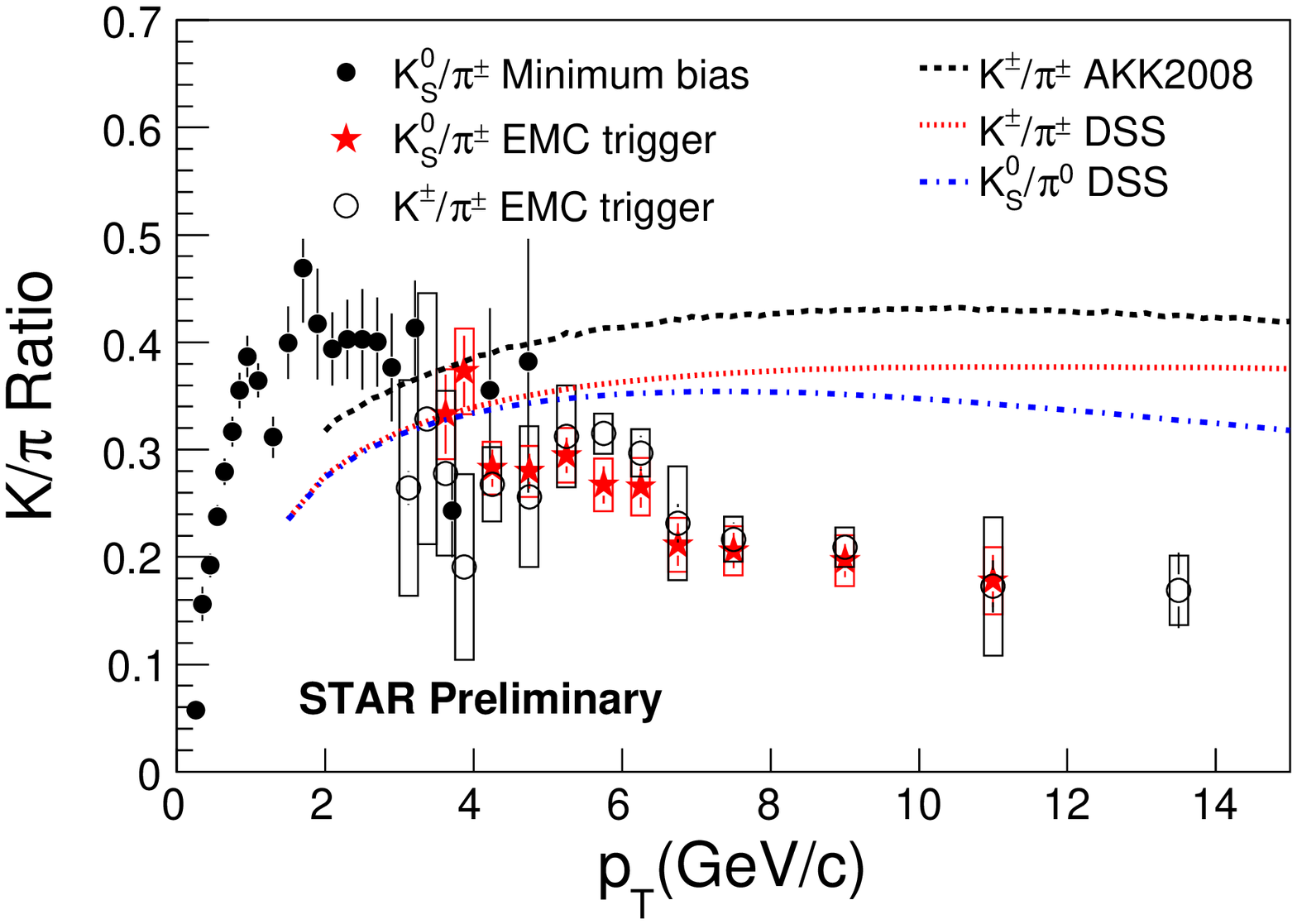}
\caption[]{Charged and neutral kaon $p_T$ spectra and K/$\pi$ ratios
in p+p collisions, compared with pQCD NLO calculations from DSS, AKK
2008. Squares and triangles are from minimum bias events, and
circles and stars are from the BEMC triggered events. The shaded
band are systematic uncertainties for $K^0_S$.} \label{ppKs}
\end{figure}

With the dE/dx information from the TPC, charged particles are
identified at $3 < p_T < 15$ GeV/$c$ at mid-rapidity, and the yields
of charged kaons can be extracted from a 8-Gaussian fit to the
inclusive positively and negatively charged particle dE/dx
distributions at given momenta \cite{ming}. Unfortunately, the
experimental dE/dx value is deviated from theoretical predictions
due to empirical parameters in theory, gas multiplication gains and
noise of the TPC electronics, and pileup in high luminosity
environment. In addition, the dominant pion yields shadowing kaon
yields will induce more uncertainties for kaon. Re-calibration
method \cite{re-cali,HotQuark} was used to improve the precision of
dE/dx in TPC here, and results in more precise yields for charged
kaons by fixing re-calibrated dE/dx peak position. To correct for
trigger enhancement, PYTHIA is used to generate events, and GEANT is
used to select different trigger events by passing different
detector thresholds as real events in STAR experiment. The trigger
enhancement factor can be calculated by the spectra in triggered
events divided by that in minimum bias events. Acceptance and
efficiency (88$\%$) are studied by Monte Carlo GEANT simulations.
Final spectra at $|\eta| < 0.5$ are shown on the left panel of
Fig.~\ref{ppKs}, which is consistent with previous published spectra
in minimum bias events.

In order to cross-check if there are trigger bias in the BEMC
triggered events, $K^{0}_{S}$ was reconstructed through $K^{0}_{S}
\rightarrow \pi^{+} + \pi^{-}$ using V0 method \cite{v0} in
different triggers, HT1 and HT2 trigger. In this analysis, one
daughter pion was required to trigger the event by depositing energy
to the BEMC tower. The trigger efficiency can be extracted by
comparing pion spectra from triggered events with pion spectra in
the minimum bias events. After trigger efficiency for triggered
daughter ($\pi$), tracking efficiency for another daughter and V0
efficiency ($\sim 50\%$) were corrected for, yields of $K^{0}_{S}$
are shown as star on the left panel of Fig.~\ref{ppKs}, and
consistent with charged kaon spectra with uncertainties. The
uncertainties include statistical uncertainties and systematical
errors which contain trigger uncertainty ($\sim10\%$), momentum
resolution ($1-20\%$), efficiency uncertainty ($5\%$) and methods of
getting $K^{0}_{S}$ yields ($\sim1\%$).  Some NLO pQCD calculations
(AKK 2008 \cite{AKK2008} and DSS) shown on left panel of
Fig.~\ref{ppKs} cannot describe our charged kaon and neutral kaon
spectra, although they show good agreements with charged pion
spectra \cite{HotQuark}. This indicates our data can provide a good
constraint for the NLO pQCD calculation.

The K/$\pi$ ratios at mid-rapidity in the BEMC triggered events from
p+p collisions as a function of $p_{T}$ are shown on right panel of
Fig.~\ref{ppKs}, and compared with published results from minimum
bias $p+p$ collisions \cite{K0pp} and some predictions from pQCD
calculations (DSS and AKK 2008). All the boxes and bars on this
figure represent systematic and statistic errors. Our measurements
are consistent with published results from minimum bias events in
p+p collisions at overlapping $p_T$ range, but lower than the
predictions from DSS and AKK 2008 at high $p_T$.

\begin{figure}
\centering
\includegraphics[scale=0.5]{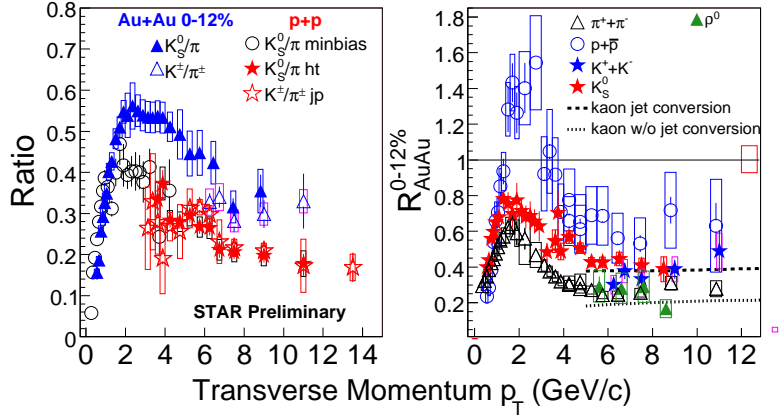}
\caption{K/$\pi$ ratios and nuclear modification factors of pion,
kaon, protons and rho in Au+Au collisions as a function of $p_T$.
The bars and boxes represent statistical and systematic
uncertainties respectively.} \label{Raa}
\end{figure}

Left panel on the Fig.~\ref{Raa} shows the K/$\pi$ ratios in p+p and
Au+Au collisions. The enhancement of K/$\pi$ ratio in Au+Au
collisions shows less suppression for kaons than pions at high
$p_T$. To further understand this phenomena, $R_{AA}$ in central
Au+Au collisions are shown on the right panel of Fig.~\ref{Raa} for
kaon, pion, proton and rho \cite{rhoQM08}. We observed that
$R_{AuAu}(K^{\pm},K^{0}_{S})$ is larger than $R_{AuAu}(\pi^{\pm})$,
which is in contradiction to the prediction from energy loss
\cite{energyLoss}. Beside, $R_{AuAu}(\pi^{\pm})$ is similar to
$R_{AuAu}(\rho)$. On the other hand, $R_{AuAu}(K^{\pm},K^{0}_{S})$
is consistent with the prediction with jet conversion in the hot
dense medium, as shown by dashed line \cite{jetConversion}. The same
factor, scaling the lowest-order QCD jet conversion rate, applied to
calculate proton $R_{AA}$ \cite{starAuAuPID,Liu} is used in this
prediction.

\section{Summary and discussion}
We report charged kaon transverse momentum spectra up to 15 GeV/$c$,
neutral kaon $p_T$ spectra up to 12 GeV/$c$ using events triggered
by the BEMC at $|\eta|<0.5$ from $p+p$ collisions at $\sqrt{s_{NN}}$
= 200 GeV. In p+p collisions, the NLO pQCD calculations cannot
describe our kaon spectra and over-predict kaon yields at high
$p_T$, which indicates our data could provide a good constraint to
FFs. At high $p_T>6$ GeV/$c$, $R_{AA}(p+\overline{p})\geq
R_{AA}(K^{0}_{S},K^{\pm})>R_{AA}(\pi^{+}+\pi^{-})\simeq
R_{AA}(\rho^{0})$. The measurements of $R_{AA}$ for kaon are
consistent with jet conversion mechanism. Other mechanisms, such as
parton splitting \cite{split} might be able to explain the
observation.

%

\end{document}